\documentclass[conference]{IEEEtran}
\IEEEoverridecommandlockouts
\usepackage{lipsum}
\usepackage{cite}
\usepackage{flushend}
\usepackage{amsmath,amssymb,amsfonts}
\usepackage{algorithmic}
\usepackage{graphicx}
\usepackage{textcomp}
\usepackage{xcolor}
\usepackage[utf8]{inputenc}
\usepackage[T1]{fontenc}
\usepackage{hyperref}
\usepackage{algorithm}
\usepackage{comment}
\def\BibTeX{{\rm B\kern-.05em{\sc i\kern-.025em b}\kern-.08em
    T\kern-.1667em\lower.7ex\hbox{E}\kern-.125emX}}

\usepackage{subcaption}
\usepackage{array}
\usepackage{caption}
\usepackage{booktabs}
\hypersetup{colorlinks=true, linkcolor=blue, filecolor=magenta, urlcolor=cyan, citecolor=blue}

\begin{document}

\title{RIS Meets O-RAN: A Practical Demonstration of Multi-user RIS Optimization through RIC}

 \author{
    \IEEEauthorblockN{Ali Fuat Şahin\IEEEauthorrefmark{1}\IEEEauthorrefmark{2}, Onur Salan\IEEEauthorrefmark{1}, Ibrahim Hokelek\IEEEauthorrefmark{1}, Ali Gorcin\IEEEauthorrefmark{1}\IEEEauthorrefmark{2} }
    \IEEEauthorblockA{\IEEEauthorrefmark{1}Communications and Signal Processing Research (HISAR) Lab, TUBITAK BILGEM, Kocaeli, Turkiye}
    \IEEEauthorblockA{\IEEEauthorrefmark{2}Faculty of Electrical and Electronics Engineering, Istanbul Technical University, Istanbul, Turkiye}
    \IEEEauthorblockA{Email: \{ali.sahin, onur.salan, ibrahim.hokelek, ali.gorcin\}@tubitak.gov.tr}
}   

\maketitle

\begin{abstract}

Open Radio Access Network (O-RAN) along with artificial intelligence, machine learning, cloud, and edge networking, and virtualization are important enablers for designing flexible and software-driven programmable wireless networks. In addition, Reconfigurable Intelligent Surfaces (RIS) represent an innovative technology to direct incoming radio signals toward desired locations by software-controlled passive reflecting antenna elements. Despite their distinctive potential, there has been limited exploration of integrating RIS with the O-RAN framework, an area that holds promise for enhancing next-generation wireless systems. This paper addresses this gap by designing and developing the RIS optimization xApps within an O-RAN-based real-time 5G environment. We perform extensive measurement experiments using an end-to-end 5G testbed including the RIS prototype in a multi-user scenario. The results demonstrate that the RIS can be effectively utilized to boost the received signal power of the selected user or provide fairness among the users.  This is a promising result demonstrating that RIS can support high-level policies in a multi-user network scenario.
\end{abstract}

\begin{IEEEkeywords}
O-RAN, OAI, RIC, xApp, reconfigurable intelligent surface, software-defined radio.
\end{IEEEkeywords}

\section{Introduction}

One of the most popular concepts in cellular mobile wireless systems is O-RAN\cite{oRAN}, where well-defined open interfaces are utilized to further promote disaggregation and softwarization of the RAN functions. Instead of monolithic structures, the disaggregation of the RAN into smaller components and their operation in the cloud platforms are expected to boost innovations from small and medium enterprises. Consequently, competitive solutions will reduce the capital and operational expenditures (i.e., CAPEX and OPEX) of the operators. The RAN Intelligent Controller (RIC) is a key innovation avenue for providing AI/ML-based resource optimization of various RAN components using the near-real time and non-real-time RIC concepts.


RIS represent a pioneering technology, where the wireless propagation environment is exploited by collectively adjusting the phases of incident signals with low-cost and low-energy elements \cite{ris}. It is an attractive technology since encoding and decoding operations are not needed, and only minimal extra energy is used for soft programming of passive reflecting elements to determine the phase of the reflected signal. RIS can be an effective tool to cope with high path loss and blockage problems, especially at higher frequencies. The integration of RIS into real-world cellular networks presents significant challenges as the high number of antenna elements may need to be re-configured as the network conditions change. Although there are many studies proposing RIS usage for next-generation cellular networks, our work specifically focuses on the integration of RIS within a multi-user O-RAN network in a realistic testbed environment including open-source 5G emulators, software-defined radios, and real RIS prototype. The literature includes studies for multi-user RIS \cite{multiRIS,multiRIS2} and real-life measurements of RIS-aided systems \cite{samedRIS}; however, they lack the O-RAN integration aspect.

In this paper, we propose a multi-user RIS optimization framework within the O-RAN architecture to extend the capabilities of the 5G network beyond the state of the art. There is a need to dynamically optimize the phase shift values of the RIS antenna elements according to time-varying network conditions and operator specified policies. The O-RAN provides flexibility for reconfiguring the RIS elements using the channel information, as the channel monitoring xApp continuously obtains wireless channel measurements from the network. The contributions of this paper include the development of novel xApps tailored for RIS in an O-RAN environment, and showcasing a practical implementation that bridges the gap between theoretical RIS benefits and real-world O-RAN applications. A multi-user RIS optimization algorithm is developed to maximize the weighted sum of the user reference signal received power (RSRP) values. The extensive measurement experiments using the RIS-assisted end-to-end 5G testbed show that the RIS can direct the RF signal toward the users resulting in higher received signal power levels. In addition to the overall performance improvements, the experiment results show that the weights in the RIS optimization algorithm can be configured to provide fairness among the users, or boost the performance of the selected user. This is a promising result demonstrating that RIS as a physical layer technology can contribute to the success of operator specified high-level policies in a multi-user network scenario. 

\begin{figure*}[h]
    \centering
    \includegraphics[width=1\textwidth]{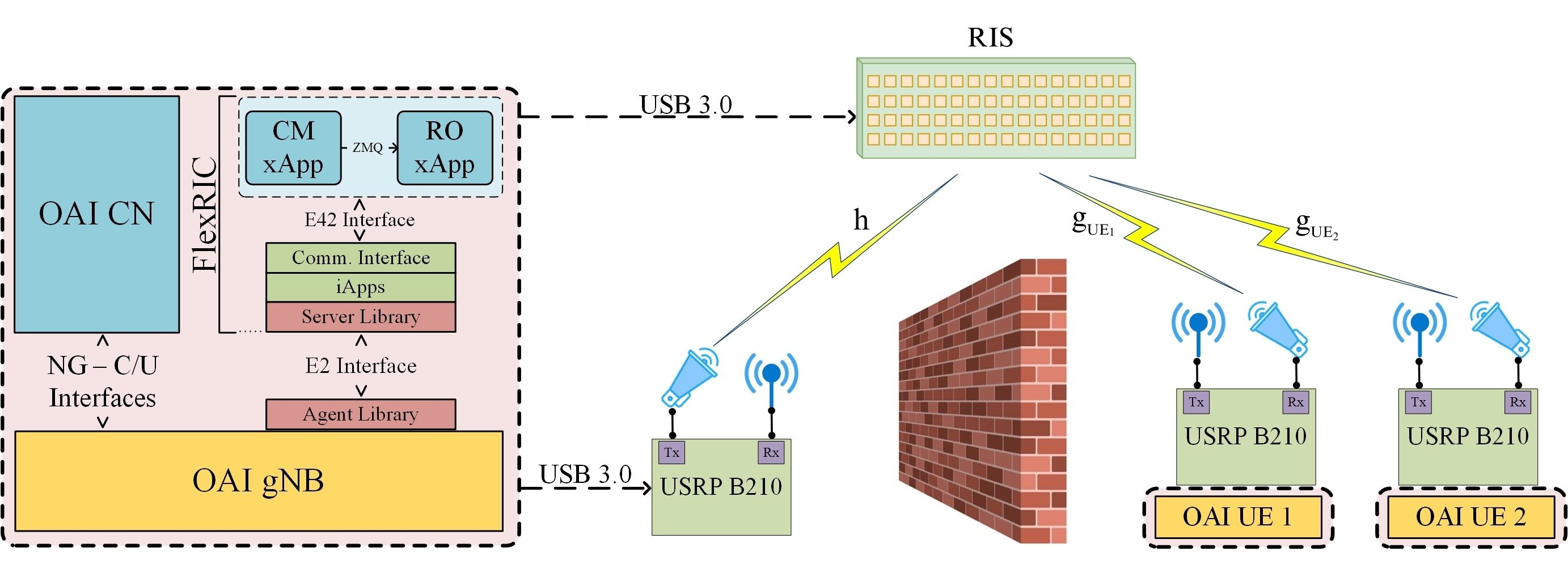}
    \caption{RIS-assisted end-to-end 5G wireless system model.}
    \label{propSysMod}
\end{figure*}

The paper is organized as follows: Section\ref{sectionII} introduces the O-RAN 5G architecture and the RIC platform. Section\ref{sectionIII} details the RIS-assisted system model, Section\ref{sectionIV} describes the multi-user RIS optimization algorithm, and Section \ref{sectionV} presents the experiment results. Finally, the paper is concluded in Section \ref{sectionVI}.
\footnote{This paper has been accepted to IEEE EuCNC, 2025.}

\section{RAN Intelligent Controller in O-RAN}
\label{sectionII}

To overcome the disadvantages of the advanced functionalities within a monolithic base station architecture, the 3GPP promotes the disaggregation through three distinct gNodeB functions starting from Release 15: Central Unit (CU), Distributed Unit (DU), and Radio Unit (RU) \cite{openRAN3GPP}. 3GPP proposes Split 2 while Split 7.2 is the most preferred architecture defined by O-RAN Alliance. Split 2 proposes the segregation of CU from DU while Split 7.2 separates DU from RU. 

The RAN Intelligent Controller (RIC) concept \cite{ric}, which enables software-based intelligent and flexible control of network resources, is an important innovation. The AI/ML-based management of limited radio resources can be performed using third-party applications such as dApps for real-time RIC, xApps for near-real time RIC, and rApps for non-real-time RIC. The near-RT RIC is responsible for managing critical latency requirements, ranging from 10 ms to 1s. The E2 interface facilitates communication between the near-RT RIC and RAN components (i.e., E2 nodes). Additionally, the E2 is also used for either periodically or trigger-based data collection from the E2 nodes to the near-RT RIC. The E2 interface utilizes two protocols, namely E2 Application Protocol (E2AP) and E2 Service Model (E2SM). The E2AP protocol is responsible for establishing the logical communication between the near-RT RIC and the E2 nodes while the E2SM protocol provides basic service types such as control, policy, report, and insert. The O-RAN Alliance has developed various E2SM service models including Key Performance Measurements (KPM), RAN Control (RC), and others \cite{oRAN}. These models provide a foundation for xApps which are micro-services designed to perform radio resource management by utilizing the E2AP and E2SM. 


The main advantage of xApps is the flexibility to apply any custom logic as needed. The only constraint for xApps is that any xApp must be defined by a software image and a descriptor. The descriptor includes essential information for the corresponding xApp such as inputs, policies, and such. Additionally, the descriptor includes the capabilities of the corresponding application. Furthermore, the near-RT RIC includes other elements such as internal messaging infrastructure, and more that have distinct responsibilities. The main objective of these elements is to provide reliable and effective management of xApps within the near-RT RIC. 

\section{RIS-Assisted E2E 5G System Model}
\label{sectionIII}

Fig. \ref{propSysMod} shows the RIS-assisted wireless communication system model, where an end-to-end 5G system consists of a core network (CN), gNodeB (gNB), and two user equipments (UEs). The channel between the transmitter and the RIS is defined as $\mathbf{h} \in \mathbb{C}^{N}$ while the channel between the RIS and the receiver is defined as $\mathbf{g} \in \mathbb{C}^{N}$. The total number of RIS elements is defined as $N = N_x \times N_y \ $ since RIS is in the form of a uniform planar array. The Line-of-Sight (LoS) channel can be defined as $h_{\text{LoS}} \in \mathbb{C}$. The complex baseband channel can be denoted as $x[k]$ while the complex Gaussian noise component at the receiver can be denoted as $n[k] \sim \mathcal{N}_{\mathbb{C}} (0, \sigma_n^2)$ where $k$ stand for discrete time instant. According to the definitions, the received signal can be given as follows:
\begin{equation}
    r[k] = \left( \mathbf{g}^H \boldsymbol{\Theta} \mathbf{h} + h_{\text{LoS}} \right) x[k] + n[k]
    \label{risReceivedSignal}
\end{equation}
where $\boldsymbol{\Theta}$ represents the RIS configuration matrix. The corresponding matrix is a diagonal matrix, defined as $\operatorname{diag} \left\{ \alpha_1 e^{j \phi_1}, \dots, \alpha_N e^{j \phi_N} \right\}$ where each diagonal component stands for the state of the corresponding RIS element. It should be noted that there is no direct LoS communication between the gNB and two UEs. Hence, the channel gain for LoS can be denoted as zero (i.e. $h_\text{LoS} = 0$). Therefore, a RIS is strategically placed to establish the LoS communication link. Following, the received signal for two users can be expressed as follows:
\begin{equation}
    r_{\text{UE}_1}[k] = \left( \mathbf{g}_{\text{UE}_1}^H \boldsymbol{\Theta} \mathbf{h} \right) x[k] + n_{\text{UE}_1}[k]
\end{equation}
\begin{equation}
    r_{\text{UE}_2}[k] = \left( \mathbf{g}_{\text{UE}_2}^H \boldsymbol{\Theta} \mathbf{h} \right) x[k] + n_{\text{UE}_2}[k]
\end{equation} 
Then, the average received power for each user in decibels relative to full scale (dBFS) is represented as follows:
\begin{equation}
    P_{\text{UE}_1} = 10 \log_{10} \left(\frac{1}{K} \sum_{k=1}^{K} \left| r_{\text{UE}_1}[k] \right|^2 \right)
    \label{Equation_UE1}
\end{equation}
\begin{equation}
    P_{\text{UE}_2} = 10 \log_{10} \left( \frac{1}{K} \sum_{k=1}^{K} \left| r_{\text{UE}_2}[k] \right|^2 \right)
    \label{Equation_UE2}
\end{equation}
where $K$ stands for the total amount of measurement samples. Furthermore, the corresponding metric in this study is defined as synchronization signal RSRP (SS-RSRP) in ETSI standards \cite{ss_rsrp}. SS-RSRP represents the linear average of power contributions of resource elements which carry the corresponding synchronization signals. As Eqs. \ref{Equation_UE1} and \ref{Equation_UE2} consider the received power of synchronization signal within a single resource element, the mathematical expression for $\text{SS-RSRP}_{i}$ of $\text{UE}_{i}$ can be calculated as follows:
\begin{equation}
    \text{SS-RSRP}_{i} = \frac{1}{N_{\mathrm{RE}}} \sum_{n=1}^{N_{\mathrm{RE}}} P_{\mathrm{{UE_{i}}}}(n)
    \label{ss_rsrp}
\end{equation}
where ${N_{\mathrm{RE}}}$ denotes the number of resource elements used in the measurement while $P_{\mathrm{{UE_{i}}}}(n)$ stands for the power contribution of $n$th resource element in dBm for each user in the network (i.e., $i  = 1,2$). From Eq. \ref{ss_rsrp}, it can be observed that the $\text{SS-RSRP}$ value is directly proportional with the end-to-end RIS channel gain, $|\mathbf{g}^H \Theta \mathbf{h}|^2$.

The RIS prototype from Greenerwave focuses transmitted signals in desired directions using software-controlled phase shifts and operates with a 1 GHz bandwidth at 5.2 GHz center frequency \cite{greenerwaveRIS}. The RIS has a total of 76 reflecting elements with adjustable reflection coefficients via PIN diodes, which configure phase shifts to either 0° or 180°. Each element has two PIN diodes for horizontal and vertical polarization, allowing four possible states per element. 

For the software component of the system, OpenAirInterface (OAI) from the OpenAirInterface Software Alliance has been deployed to emulate the end-to-end 5G standalone (SA) system \cite{oai}.  The OAI base station emulation software runs on a COTS computer, which is connected to USRP B210 software-defined radio (SDR), to emulate the gNB functionality. For the OAI emulation parameters, the NR band 79, which is located under FR1, is selected with the carrier frequency of 4.9 GHz. Note that the NR band 79 operates in Time-Division Duplexing (TDD) mode which uses the same frequency range for both downlink and uplink communication. Furthermore, a subcarrier spacing of 15 kHz is employed, and each user in the network is allocated 20 MHz of bandwidth. The OAI UE emulation software runs on a COTS computer, which is connected to USRP B210 SDR, to emulate the UE functionality. The OAI CN software is utilized to generate an end-to-end 5G standalone (SA) system. For the downlink communication, horn antennas are deployed at the transmitter side of the gNB and the receiver side of the UEs. For the uplink communication, omnidirectional antennas are utilized in the opposite direction. Note that this study focuses only on downlink communication, where the RIS does not cause any interference among UEs as the downlink signal needs to be received by all UEs connected to the network. 

In this study, together with the OAI, we utilize the open-source FlexRIC software developed by the OAI community as the near-RT RIC platform to develop two novel xApps, the first one for wireless channel monitoring and the second one for RIS phase shift optimization \cite{flexric}. The data communication between two xApps is implemented through a zero message queue (ZMQ) interface since the RIS Optimization xApp needs to receive the performance measurements after reconfiguring the phases of the RIS antenna elements through the USB interface. 

\section{Multi-User RIS Optimization Algorithm}
\label{sectionIV}

In this study, we propose a multi-user RIS optimization algorithm to maximize the weighted sum of the user SS-RSRP values as follows:
\begin{equation}
    \text{RSRP}_{WS} = \sum_{i=1}^{N_{UE}} w_i * (\text{SS-RSRP}_i) 
    \label{RSRP_WS}
\end{equation}
where $N_{UE}$ and $w_i$ denote the number of users and the weight of the user $i$. Note that the mathematical expression for $\text{SS-RSRP}_i$ is given in Eq. \ref{ss_rsrp}. The user weights can be determined by three different approaches according to the network administrator's specified policies. In the first method, users are assigned equal weights without considering their channel qualities. The second and third methods determine the weights by utilizing the users' channel quality information. In the second method, the weights are calculated using the proportional fair approach. The third method optimizes the RIS phase shifts by further improving the RSRP value of the strong user while the weak user has a lower priority using a lower weight value. 

For the optimization of the RIS, the reference algorithm proposed in \cite{hisarRIS,samedRIS} has been modified to accommodate multiple UEs. The pseudo-code of the algorithm is given in Algorithm \ref{risOpt}. The phase shifts of the RIS elements are represented by the $States$ vector which is initially set to 0. This $States$ vector corresponds to the RIS configuration, where all PIN diodes are set to off state. Each RIS element can take $N_{state}$ different states (i.e., $N_{state}$ distinct phase shifts). The corresponding element of $States$ is reconfigured using the phase shift of the selected element that maximizes the weighted sum of the RSRP value (i.e., $RSRP_{WS}$). This process is repeated sequentially for all elements. It should be noted that, due to constraints imposed by the near-RT RIC, this algorithm operates with a step duration of 10 ms.

\begin{algorithm}
\caption{RIS Optimization through RIC Algorithm}
\begin{algorithmic}[1]
\renewcommand{\algorithmicrequire}{\textbf{Input:}}
\renewcommand{\algorithmicensure}{\textbf{Output:}}
\ENSURE $States$
\STATE initialize $States$ to $0^\circ$ phase shift
\STATE collect $\text{SS-RSRP}_i$ and calculate $w_i$ for user $i$ 
\STATE calculate $\text{RSRP}_{WS,max}$
\STATE $ElementsRIS \gets shuffle \left[1, N_x \times N_y \right]$
\FOR{each $i \in ElementsRIS$}
    \FOR{each $j \in [1, N_{state}]$}
        \STATE configure $i$th element with state $j$
        \STATE collect $\text{SS-RSRP}_i$ and calculate $w_i$ for user $i$
        \STATE calculate $\text{RSRP}_{WS}$
        \IF{$\text{RSRP}_{WS} > \text{RSRP}_{WS,max}$}
            \STATE update $i$th element of $States$ with state $j$
            \STATE $\text{RSRP}_{WS,max} \gets \text{RSRP}_{WS}$
        \ENDIF
    \ENDFOR
\ENDFOR
\RETURN $States$
\end{algorithmic}
\label{risOpt}
\end{algorithm}

\begin{figure*}[h]
    \centering
    \includegraphics[width=0.75\linewidth, trim={0 0 0 13cm},clip]{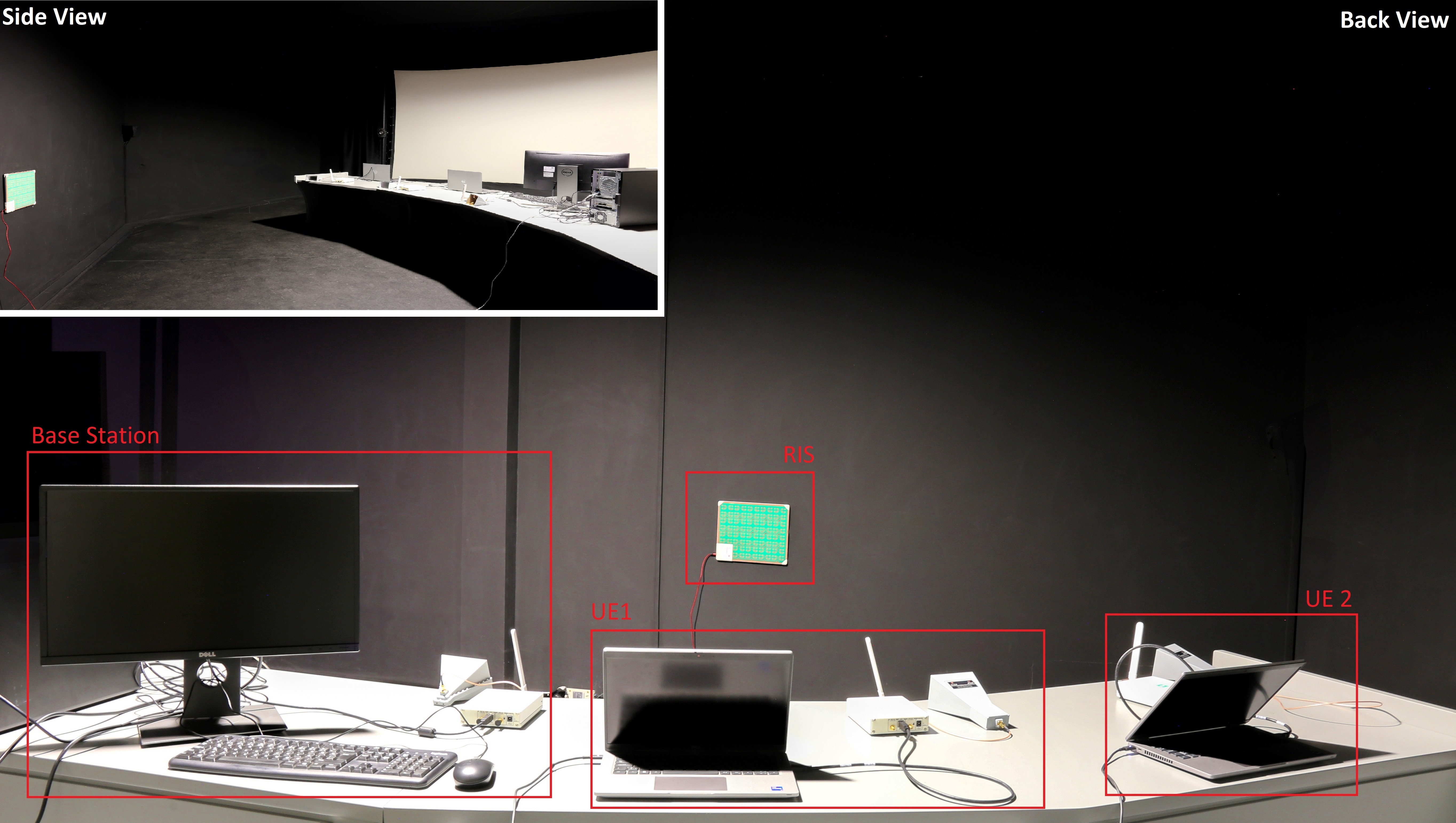}
    \caption{Experimental Setup.}
    \label{expSetup}
\end{figure*}

\textit{Method 1}: The RIS phase shifts are optimized using Algorithm 1, where the weights of the users are set to equal values. The expected outcome is to have a similar performance improvement for all users as their weights are the same. The weights for Method 1 are calculated as follows: 

\begin{equation}
    w_i = 1 / N_{UE}, \quad \forall i=1,\cdots,N_{UE}
    \label{w_RR}
\end{equation}

\textit{Method 2}: This method aims to demonstrate how the RIS phase shift optimization should be performed when the users have distinct channel qualities (e.g., one user has significantly worse channel quality compared to the other user). This approach provides a higher weight to the user with a poor channel condition, and hence the expected outcome is to have a higher performance improvement for this disadvantaged user. In this case, the weight calculation method is inspired by the Proportional Fair scheduling algorithm in 5G NR \cite{schedulingAll}. 

Firstly, the transport block size for the $i$th user ($TBS_i$) is computed as described in TS 38.214 \cite{ts38214}. The next step is to measure the throughput value of the downlink shared channel for the $i$th user ($R_{\text{DLSCH}, i}$). Finally, the weights of the corresponding users are calculated as follows:

\begin{equation}
    w_i = TBS_i / \text{R}_{\text{DLSCH},i} \quad \forall i=1,\cdots,N_{UE}
    \label{w_PF}
\end{equation}

\textit{Method 3}: This method sets the weights of the users proportional to their channel quality indicator (CQI) values. The main motivation is to optimize the RIS phase shifts such that the highest performance improvement is achieved for the user having the highest channel quality. This is expected to yield the highest system throughput similar to the Best CQI scheduling algorithm \cite{schedulingAll}. The weights are calculated using the CQI values of the users as follows: 

\begin{equation}
    w_i = \frac{CQI_i}{\sum_{i=k}^{N_{UE}}CQI_k} , \quad \forall i=1,\cdots,N_{UE}, \quad CQI = 1,\cdots,15
    \label{w_BestCQI}
\end{equation}

\section{Experiment Results}
\label{sectionV}

In this section, we present the experimental setup and the results of the measurement experiments using the end-to-end 5G testbed including the RIS prototype. Firstly, a 5G CN is started for the management and control of the 5G network, and then a monolithic gNB is employed. The RIC is connected to the gNB through the E2 interface. Subsequently, two UEs are started, and the registration of the gNB and two UEs to the network is verified through the 5G CN log files. Afterward, the channel monitoring xApp is activated to obtain channel measurements of both UEs, where the measurement parameters are defined under the 3GPP standards known as key performance measurements. The channel monitoring (CM) xApp gets the channel information of the users and sends this information to the RIS Optimization (RO) xApp through the ZMQ interface. 

\begin{figure}[h]
    \centering
    \includegraphics[width=0.5\textwidth]{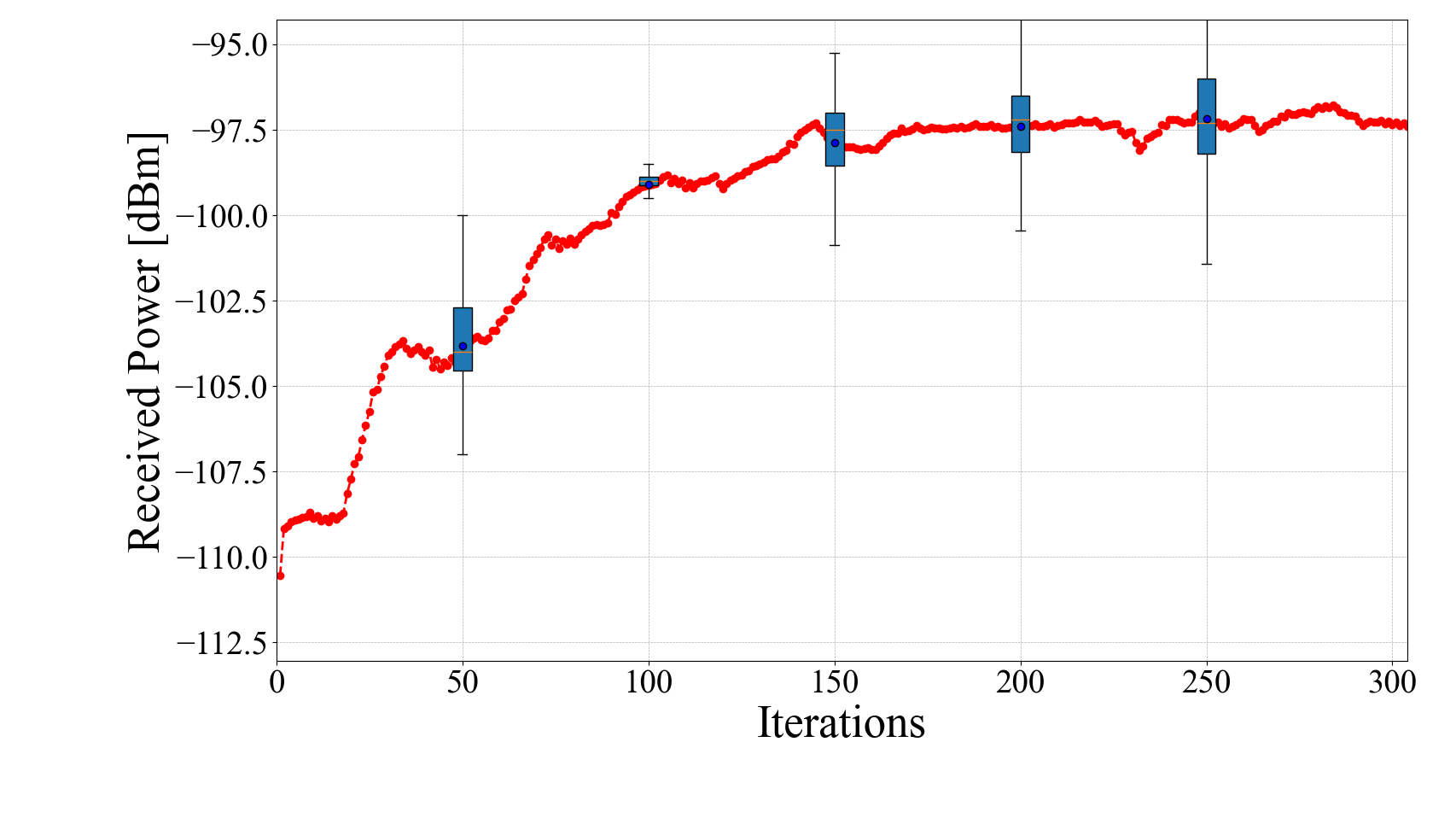}
    \caption{RIS Optimization for a single UE.}
    \label{Res1}
\end{figure}   

\begin{figure*}[t]
    \centering
    \begin{subfigure}[b]{0.49\textwidth}
        \centering
        \includegraphics[width=\textwidth]{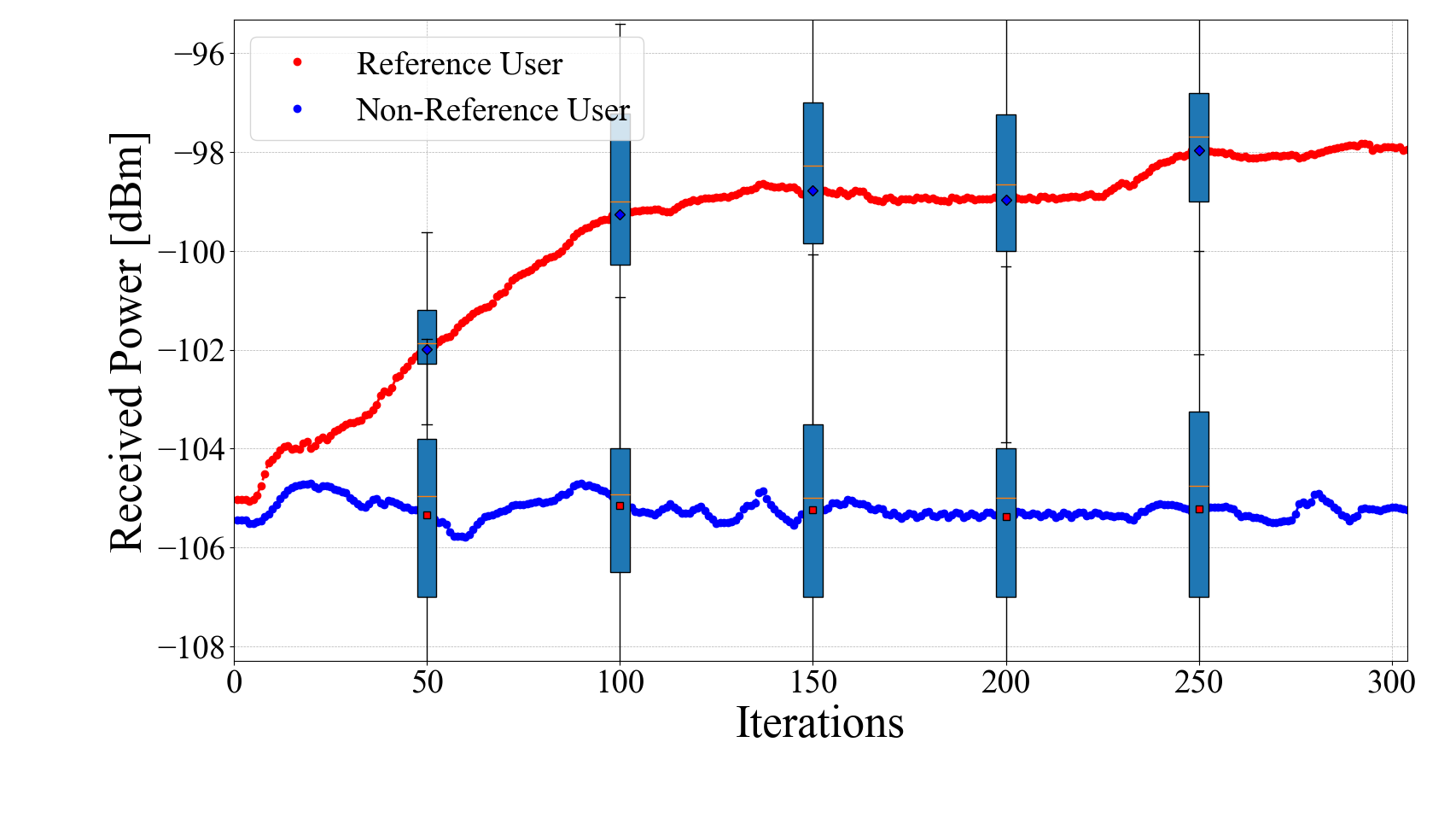}
        \caption{A single reference UE.}
        \label{Res2}
    \end{subfigure}
    \hfill
    \begin{subfigure}[b]{0.49\textwidth}
        \centering
        \includegraphics[width=\textwidth]{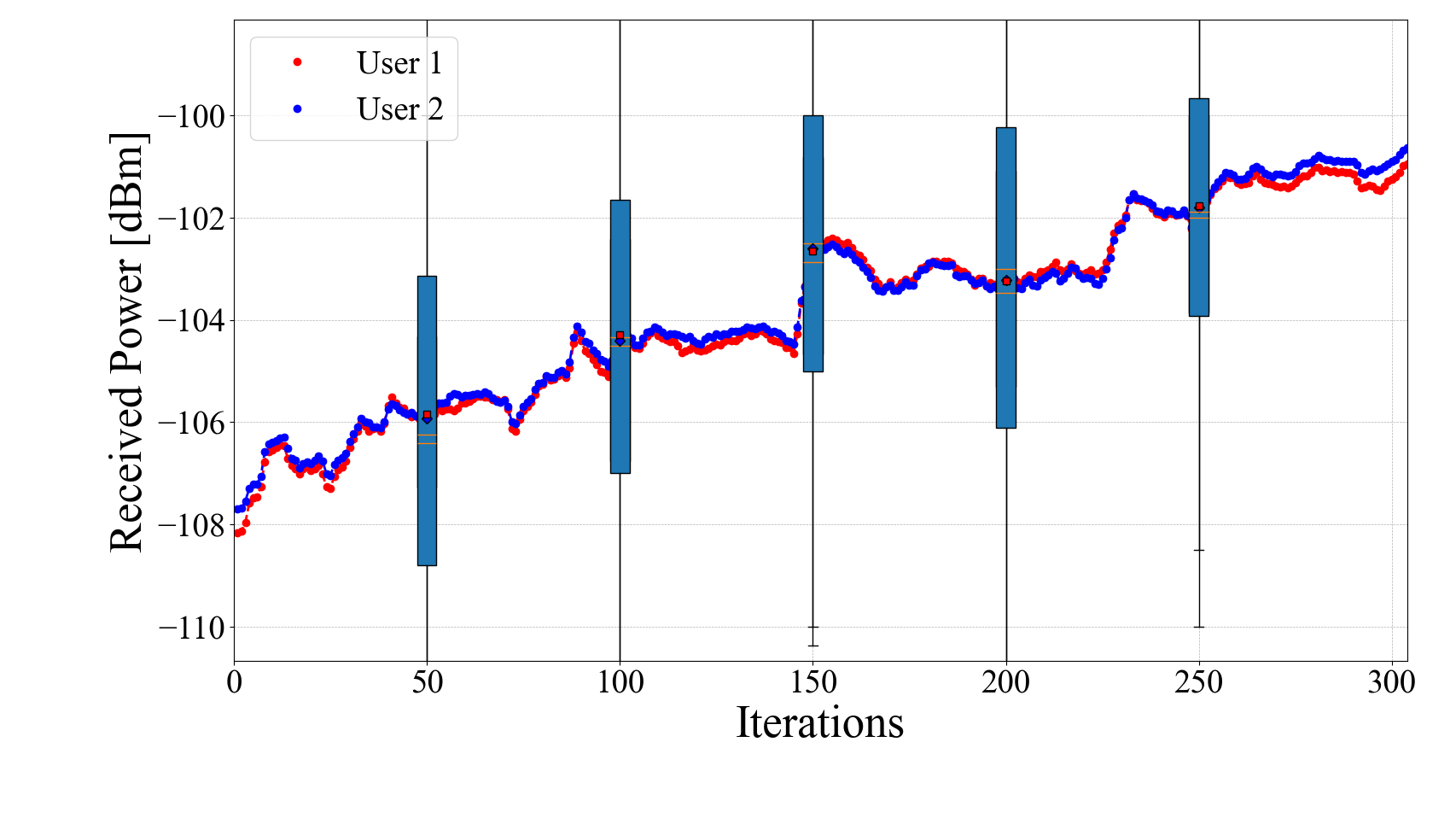}
        \caption{Method 1.}
        \label{Res3}
    \end{subfigure}
    \vskip\baselineskip
    \begin{subfigure}[b]{0.49\textwidth}
        \centering
        \includegraphics[width=\textwidth]{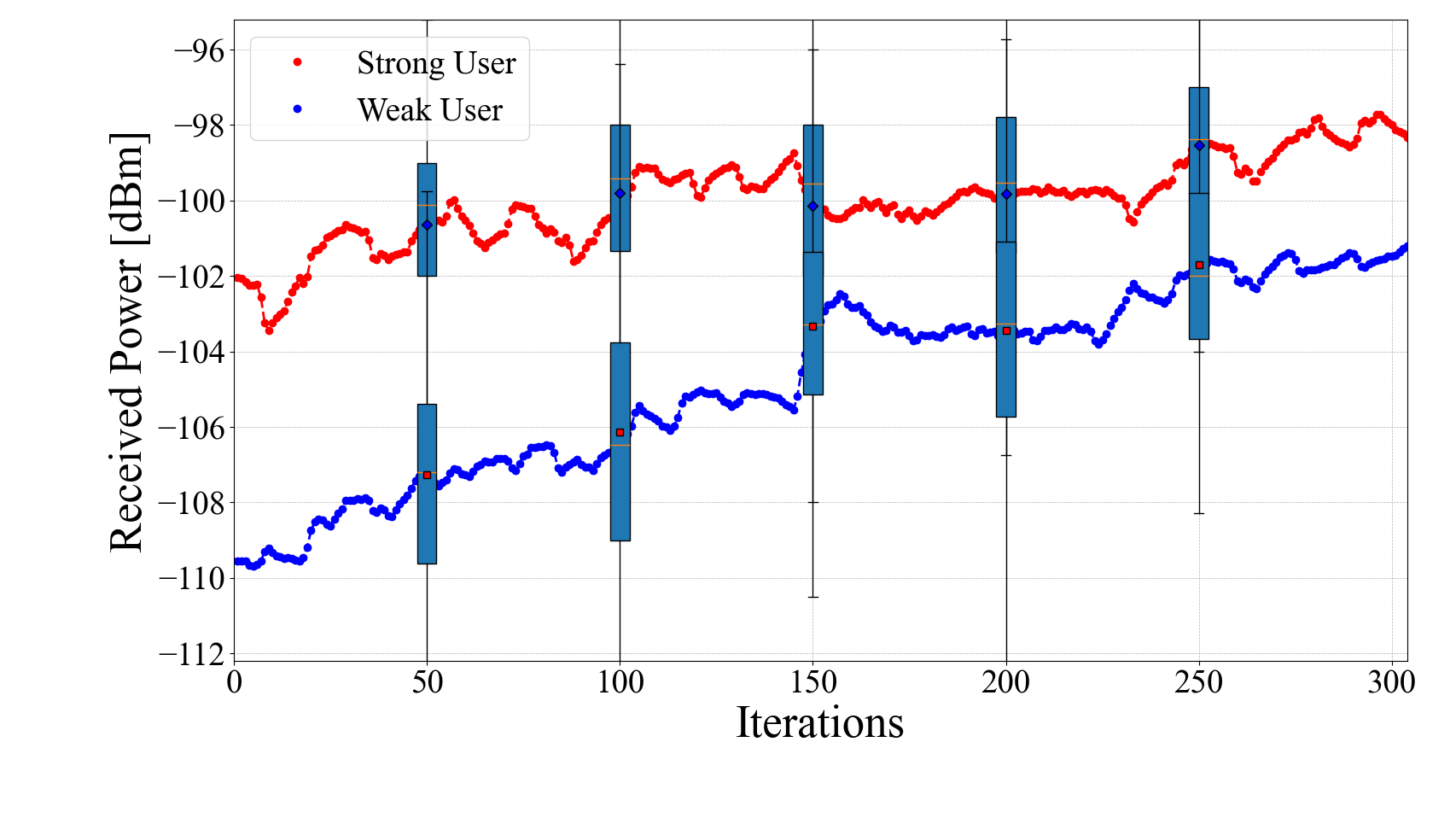}
        \caption{Method 2.}
        \label{Res4}
    \end{subfigure}
    \hfill
    \begin{subfigure}[b]{0.49\textwidth}
        \centering
        \includegraphics[width=\textwidth]{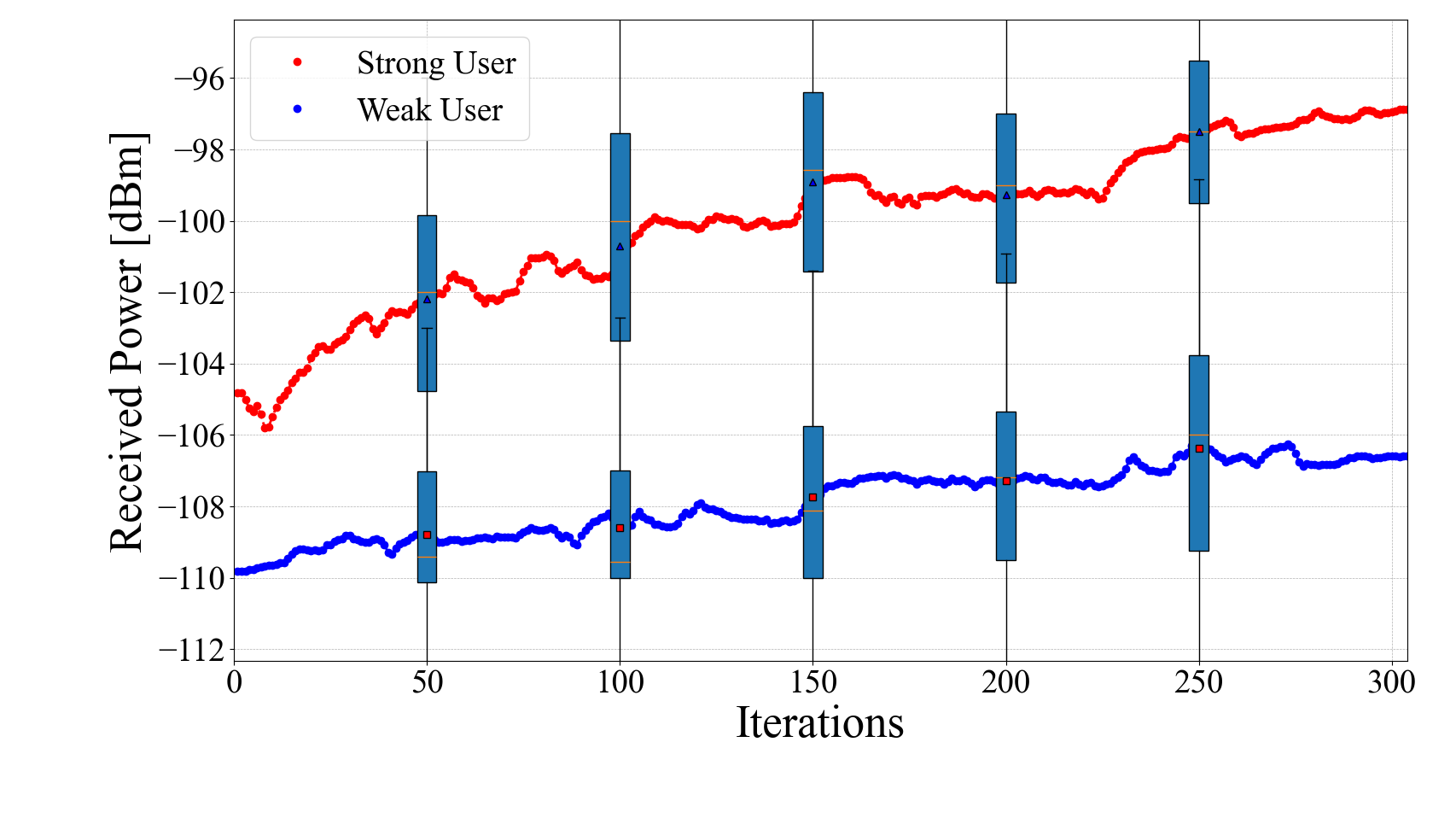}
        \caption{Method 3.}
        \label{Res5}
    \end{subfigure}
    \caption{The Comparison of RIS Optimization Methods.}
    \label{allRes}
\end{figure*}

The experimental setup is shown in Fig. \ref{expSetup}, where 5G CN, 5G RAN, FlexRIC, and the corresponding xApps are deployed on a single computer. A USRP B210 software-defined radio (SDR) is used for RF signal transmission and reception at the base station (gNB), with a horn antenna for the transmitter and an omnidirectional antenna for the receiver. For both UEs, two separate computers along with two USRP B210s are utilized. In contrast to the gNB, both B210s are equipped with horn antennas at the receiver side and omnidirectional antennas at the transmitter side. It is important to note that hardware and software components are identical for both UEs. The Greenerwave RIS prototype includes 76 RIS elements, each with four states, resulting in a total of 304 iterations for a single loop of the RIS optimization algorithm described in Algorithm \ref{risOpt}. For the performance results, the enhancement of the initial SS-RSRP values is reported for each scenario. For each case, we performed 300 experiments, and the mean RSRP values of the UEs are reported in the figures.  

In the first set of experiments, there is only a single UE in the 5G network, and the RIS optimization is performed by considering this particular UE performance. The objective of this case is to simply understand the impact of RIS configuration within a 5G network. The results in Fig. \ref{Res1} show that the SS-RSRP values are around -110 dBm. Following the RIS optimization, the SS-RSRP values increase to approximately -97.5 dBm.  The results correspond to approximately 13 dBm performance improvement, indicating that the RIS can effectively enhance the UE performance in a 5G system.

The second case includes two UEs within the 5G network, where one of the UE is selected as the reference UE such that the RIS phase shift optimization is performed for this reference UE without taking the second UE into account. The results in Fig. \ref{Res2} indicate that the reference UE experiences a significant increase in its RSRP values while the second UE (i.e., the non-reference user) does not experience any notable change in its RSRP values. Quantitatively, the reference user experiences an RSRP enhancement of approximately 7 dBm whereas the non-reference user observes an RSRP enhancement of only about 0.2 dBm. The results indicate that the RIS can be strategically optimized to enhance the performance of the selected user. 

The third case includes two UEs within the 5G system when the RIS phase shift optimization is performed using the weight calculation by Method 1. The corresponding results are shown in Fig. \ref{Res3}, where the RIS provides uniform enhancements for both UEs. Specifically, the first UE experiences an RSRP improvement of approximately 7.2 dBm while the second UE experiences an RSRP improvement of around 7 dBm. Both users experience similar performance increases through RIS optimization. Note that all measurements were taken under identical conditions, including distance, angles, and other factors. When the results are compared with the single reference case shown in Fig. \ref{Res3}, it is clear that the performance improvements are not attributable to physical layer parameters such as the users' positions or distances. Considering this, it can be suggested that approximately equal performance improvements for both users are due to Method 1.

The fourth case examines a scenario, where two UEs experience different quality of services (QoS). The proposed algorithm aims to improve the performance of both users' connections, with a higher priority given to the user with the lower QoS, referred to as the "weak user". Conversely, the user with higher QoS values is designated as the "strong user". The results of the proposed algorithm are given in Fig. \ref{Res4}. As expected, both users experienced significant improvements in channel conditions. The weak user can observe an RSRP improvement of around 8.4 dBm while the strong user can only observe an RSRP improvement of around 4 dBm. These results indicate that the proposed algorithm works as intended. It should be noted that, unlike previous scenarios, the proposed algorithm dynamically assigns weights to each user. In this case, when a user's performance improves, their assigned weight is reduced to achieve more fair results. Around the 140th iteration, the strong user's performance improves steadily while the weight decreases accordingly. Consequently, around the 150th iteration, the RIS focuses on improving the weaker user's performance due to their higher weights. A similar situation is observed around the 225th iteration where the weak user is assigned higher weights while the strong user is assigned with lower weights. From these results, it can be observed that the proposed algorithm is well-suited for scenarios, where fairness is an essential metric for the RIS deployment within the 5G network.

Similar to the fourth case, the fifth case involves two UEs experiencing different QoS. The proposed algorithm for the fifth case aims to improve the service quality of the strong user further than the weak user. As illustrated in Fig. \ref{Res5}, the strong user begins with RSRP values around -105 dBm and experiences an improvement of nearly 8 dBm, while the weak user experiences an RSRP improvement of around 3.5 dBm. In the proposed algorithm, the user weights are dynamically assigned, akin to the previous case. Since the objective is to improve the performance of the strong user relative to the weak one, it is expected that the performance gap between the two users widens as the number of iterations increases. The results confirm that the performance improvement of the strong user is greater than that of the weak user as the iterations progress. On a final note, it can be concluded that the proposed algorithm is well-suited for scenarios, where the RIS is employed to maximize the overall system performance in the presence of multiple users.

\section{Conclusion}
\label{sectionVI}

In this study, a RIS-assisted 5G system has been deployed together with an open-source RIC platform. Two xApps are developed, the first one for monitoring the users' wireless channel qualities and the second one for performing the RIS phase shift optimization. We propose multi-user RIS optimization algorithms to influence how RIS helps improving the user received signal power values. The measurement experiments using the real RIS prototype in a multi-user indoor scenario demonstrate that the RIS can be effectively used to boost the selected users' RSRP values or get the RSRP values of the UEs closer to each other. The weight selection method can be configured by the network operator to impose high level network policies. While many of the existing studies focus on how to optimize the RIS through the physical layer parameters, our work provides a novel perspective that incorporates high level network policies within the O-RAN framework. The future work will implement AI/ML techniques within the RIS optimization xApps. 

\bibliographystyle{IEEEtran}
\bibliography{refer}

\end{document}